\documentclass{aa}
\usepackage{graphicx}

\newcommand{\HI}{{\ion{H}{1}}}

\newcommand{\kms}{$\,$km$\,$s$^{-1}$}

\newcommand{\mJybeam}{mJy beam$^{-1}$}
\newcommand{\msun}{{$M_\odot$}}

\def\HI{H{\,\small I}}

\def\emph#1{{\sl #1}}
\newcommand{\ltsima} {$\; \buildrel < \over \sim \;$}
\newcommand{\gtsima} {$\; \buildrel > \over \sim \;$}
\newcommand{\lta} {\lower.5ex\hbox{\ltsima}}
\newcommand{\gta} {\lower.5ex\hbox{\gtsima}}

\input{psfig}

\begin{document}

\title{The unfriendly ISM in the \\ radio galaxy 
4C~12.50 (PKS 1345+12)}

\titlerunning{VLBI observations of neutral hydrogen in 4C~12.50}
\authorrunning{Morganti et al.}

\author{R. Morganti\inst{1}, T.A. Oosterloo\inst{1}, C.N. 
Tadhunter\inst{2},  R. Vermeulen\inst{1}, Y.M. Pihlstr\"om\inst{3}, \\ 
G. van Moorsel\inst{3}, K.A. Wills\inst{2}
}

\offprints{morganti@astron.nl}

\institute{Netherlands Foundation for Research in Astronomy, Postbus 2,
NL-7990 AA, Dwingeloo, The Netherlands
\and
Dep. Physics and Astronomy,
University of Sheffield, Sheffield, S7 3RH, UK
\and
National Radio Astronomy Observatory, Socorro,
             NM 87801, USA}  

\date{Received ...; accepted ...}

\abstract{The radio source 4C~12.50 has often been  suggested to be a prime
candidate for the link between ultraluminous infrared galaxies and
young radio galaxies.  A VLBI study of the neutral hydrogen in the
nuclear regions of this object shows that most of the gas detected
close to the systemic velocity is associated with an off-nuclear cloud
($\sim 50$ to 100 pc from the radio core) with a column density of
$\sim 10^{22}\ T_{\rm spin}/(100\ {\rm K}$) cm$^{-2}$ and an \HI\ mass
of a few times $10^5$ to $10^6$ \msun. We consider a number of
possibilities to explain the results. In particular, we discus the
possibility that this cloud indicates the presence of a rich and
clumpy interstellar medium in the centre, likely left over from the
merger that triggered the activity and that this medium influences the
growth of the radio source.  The location of the cloud -- at the edge
of the northern radio jet/lobe -- suggests that the radio jet might be
interacting with a gas cloud.  This interaction could be responsible
for bending the young radio jet. The velocity profile of the gas is
relatively broad ($\sim 150$ \kms) and we interpret this as
kinematical evidence for interaction of the radio plasma with the
cloud.  We also consider the model where the cloud is part of a
broader circumnuclear structure. Only a limited
region of this structure would have sufficient background radio brightness
and large enough column depth in neutral gas to obtain detectable \HI\
absorption against the counterjet.\\ The VLBI study of the neutral
hydrogen in 4C~12.50 suggests that \HI\ detected near the systemic
velocity (as it is often the case in radio galaxies) may not 
necessarily be connected
with a circumnuclear disk
or torus (as is very often assumed) but instead could be a tracer of
the large-scale medium that surrounds the active nucleus and that may
influence the growth of the young radio source.
}
\maketitle

\section{Introduction}

The onset of radio activity is often thought to be related to an accretion or
merger event in the parent galaxy. As the result of such an event, the galaxy
harbouring the radio source is likely to have quite a rich interstellar medium
(ISM) at least during the initial phase of the AGN. The study of this ISM has
a number of interesting aspects.  Apart from being a possible source of fuel
for the central activity, this ISM is likely to influence the growth and
evolution of the radio source as the jets propagate through the ISM on their
way out of the galaxy. Secondly, the possible interaction between the radio
plasma and the ISM is likely to produce a feedback mechanism that can put
substantial amounts of energy into the ISM. This can further influence the
evolution of the galaxy as a whole and may even be a factor that regulates the
growth of the central black hole (e.g.\ Silk \& Rees 1998).

There is a consensus now that radio galaxies classified as ``Compact
Steep Spectrum" (CSS) or ``GigaHertz Peaked Spectra" (GPS) can be
identified as objects in their initial phase of radio activity.  These
radio sources are defined as small (sub-galactic, i.e.\ \lta 10 kpc),
with steep or convex radio spectra  and
are very often double lobed, indeed like a mini-radio source (see
O'Dea 1998 for a review).  Years of observations and discussions have
led to the conclusion that these objects are not small because the
radio jets are trapped in the surrounding gaseous material, but
because they are young (see e.g. Owsianik, Conway \& Polatidis 1998).
However, reaching this conclusion has not been straightforward as
there are indications that in these objects interactions between the
radio plasma and the ISM do also occur. The ISM may not be trapping
the radio source, but it may influence its evolution. 

Evidence for some struggle of the radio jet in emerging from the nuclear
regions of CSS/GPS comes, for example, from detailed optical studies of a
number of cases.  Outflows of ionized gas are now unambiguously detected in
the two young radio sources PKS~1549--79 and 4C~12.50 (Tadhunter et al.\ 2001,
Holt, Tadhunter \& Morganti\ 2003).  Furthermore, broad optical
(forbidden) emission lines are typically observed in CSS/GPS sources,
indicating the presence of gas with disturbed kinematics as a result of an
interaction of the radio plasma with the ISM (Gelderman \& Whittle 1994).

In the scenario considered above of radio galaxies originating by
merger/interaction events, these characteristics can be explained
quite naturally.  These young radio sources would be still surrounded
by (at least some) material left over from the merger event that triggered the
radio source. The study of this medium is, therefore particularly
important if we are to understand how the galaxy evolves.

Unfortunately, optical studies of the ionized gas lack the very high spatial
resolution that is required to directly infer the details of how the gas is
distributed around the nuclear regions of the sources.  As this medium is
known to be multi-phased -- from ionized to neutral to molecular -- a
(complementary) way to study it is through high resolution (VLBI) radio
(21-cm) observations of the neutral hydrogen.  In this letter we present the
results of such a study of the GPS source 4C~12.50.  This is a particularly
interesting object as it is a prime candidate for the link between
ultraluminous infrared galaxies (ULIGs, Sanders \& Mirabel 1996) and radio
galaxies.  Indeed, the ISM of this radio galaxy is very rich as it is the most
far-IR luminous radio galaxy (L$_{\rm {IR}}=1.7 \times 10^{12}$ L$_\odot$) and
has a high molecular gas mass ($\sim 10^{10}$ M$_\odot$, Evans et al.\ 1999)
and a young stellar population component (Grandi 1977; Tadhunter et al. 
2004).  The substantial contribution found from this stellar component
(estimated to have an age between 0.5 and 1.5 Gyr) implies a major merger as
origin of this system.  4C~12.50 is therefore a good candidate in which to
study the effects of a rich ISM.  The radio source has a size of $\sim 0.15$
arcsec ($\sim 325$ pc \footnote{For $H_\circ = 71$ \kms Mpc$^{-1}$,
$\Omega_{\rm M} = 0.27$ and $\Omega_{\rm vac} = 0.73$, resulting in a linear
to angular scale ratio of 2.165 kpc arcsec$^{-1}$. Derived using E.L.\
Wright's cosmology calculator at  
http://www.astro.ucla.edu/wright/CosmoCalc.html)} and has all the
characteristics of young radio sources (with age $<< 10^7$ yr). 

\begin{figure*}
\centerline{\psfig{figure=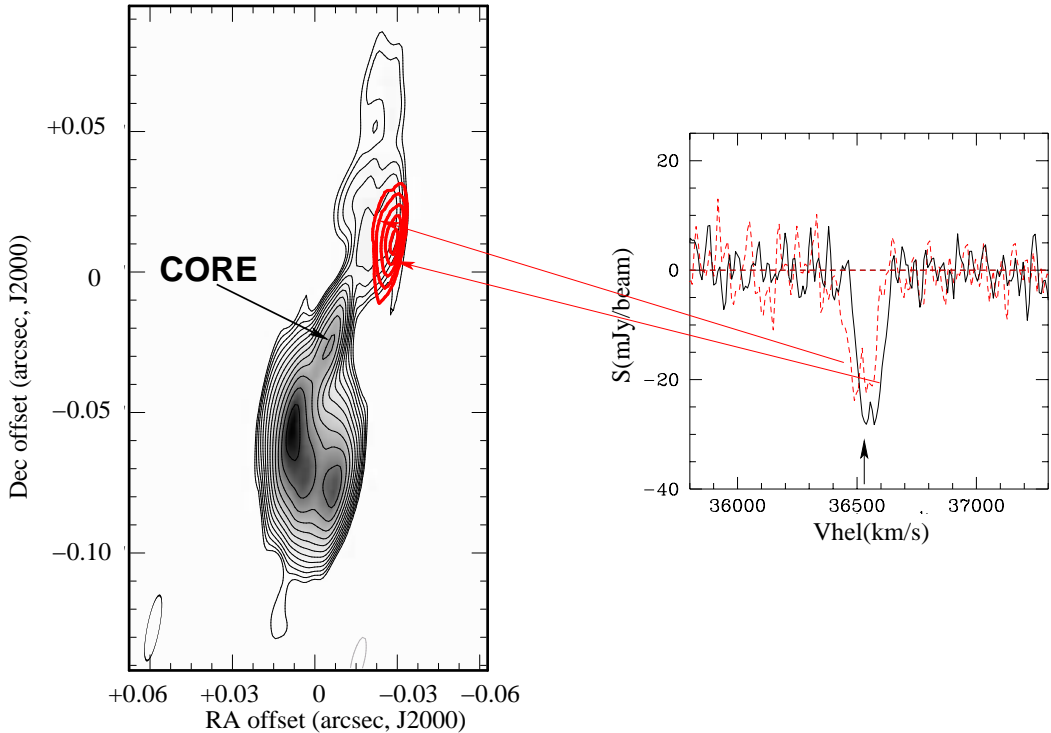,angle=0,width=16cm}}
\caption{Continuum image (grey scale and thin contours; negative contour in
light grey) of 4C~12.50 with superimposed the total intensity of the \HI\
absorption (thick contours). The absorption profiles at the two sides of the
\HI\ cloud are also shown.  The position of the radio core as derived by
Stanghellini et al. (1997) and confirmed by Lister et al. (2003) is also
indicated. The contour levels for the continuum image are: -5, 5 \mJybeam\ to
800 \mJybeam\ in steps of factor 1.5.
The contour levels for the total intensity of the \HI\ absorption are -2.5,
-2.0, -1.5, -1.0, -0.5 Jy beam$^{-1}$ \kms.} 
\end{figure*}

\HI\ absorption was detected in 4C~12.50 by Mirabel (1989) using the  
Arecibo telescope, showing a relatively deep and narrow component ($\tau \sim
0.01$ and FWHM $\sim 200$ \kms) located close to the systemic velocity as well
as a shallower and broader component ($\tau \sim 0.002$, FWHM $\sim 700$
\kms). Recent observations done with the new broad-band system at the WSRT
have given further insight the characteristics of the absorption. These
observations show that the broad absorption is much wider than measured before
by Mirabel (1989), covering now a range in velocities of about 2000 \kms; most
of this \HI\ absorption is blueshifted (see Morganti et al.\ 2003a and the
profile shown in Fig.\ 2).

Holt et al.\ (2003) have found complex optical emission line profiles
at the position of the nucleus with line width up to 2000 \kms\ and
blueshifted $\sim 2000$ \kms\ with respect to the halo of the galaxy
and the deep, narrow component of the \HI\ absorption. These findings
have been interpreted as material in outflow.  They support the idea
that 4C~12.50 is a young radio source with nuclear regions enshrouded
in a dense cocoon of gas and dust and where the radio jets are
expanding through this cocoon, sweeping material out from the nuclear
regions.  In light of the considerations mentioned above, 4C12.50
appears to be a good candidate for further study of the influence of the
ISM on the evolution of a young radio source. This was, therefore,
the goal of our VLBI study of the \HI\ in this radio source.

\section{VLBI observations}

The observations were done using a global VLBI network. This included the
VLBA\footnote{The National Radio Astronomy Observatory is operated by
Associated Universities, Inc., under cooperative agreement with the National
Science Foundation}, one antenna of the VLA and the three EVN\footnote{The European VLBI
Network is a joint facility of European, Chinese, South African and other
radio astronomy institutes funded by their national research councils.}
telescopes (Effelsberg,
Westerbork, Onsala) that can reach the relatively low frequency of the
redshifted \HI\ in 4C~12.50 (1266~MHz) with their L-band receivers.  The
observations were carried out on 24 February 2001. 4C~12.50 was observed
for about 9h. Two calibrators, 3C273 and 4C39.25, were observed, interleaved
with the main target.

When these observations were planned, we were not yet aware of the
very large width of the \HI\ absorption that was later detected by the
WSRT.  Therefore, in the VLBI observations we have used an 8~MHz band
centred on the frequency of the deep and narrow absorption (1266~MHz).
This setup is, therefore, not ideal for the study of the broad
component. Thus, in this letter, we will only concentrate on the
results from the study of the narrow component.  A total number of 256
channels were used given a velocity resolution of about 8 \kms\ before
Hanning smoothing.  The data were correlated at the JIVE correlator in
Dwingeloo.  The initial calibration used the automatic pipeline
developed at JIVE (Reynolds et al.\ 2002).  The data reduction,
including bandpass calibration, was done using the Difmap and Miriad
packages. A few cycles of phase-only self-calibration were done until 
the structure of the source appeared settled. A last iteration with  amplitude
self-calibration was then performed.

A line data cube was made using natural weighting after subtracting
the continuum emission from the $uv$-data using the channels that did
not contain the narrow \HI\ absorption. This approach was chosen because
the 8 MHz observing band does not allow to determine the properties of
the broader absorption. The noise per channel is $\sim 1.3$
\mJybeam\ after Hanning smoothing  and the restoring beam size is $24
\times 4$ mas (p.a.\ = $-11^{\circ}$).  The reason for choosing this low 
resolution is that it allowed the detection of the faint, extended part of the
continuum source that turned out to be critical (see below).

A continuum image (at the same spatial resolution) was made using the
line-free channels. The noise is $\sim 1$ \mJybeam\ and the resulting image is
shown in Fig.~1.  The well-known, heavily distorted, structure of this source
is clearly evident (see also Stanghellini et al.\ 1997, Xiang et al.\ 2002,
Lister et al.\ 2003).  The location of the core, as derived by Stanghellini et
al.\ (1997) is indicated. The peak of the continuum is 809 \mJybeam\ and the
total flux density is 5.17 Jy, comparable to previous measurements of the
total flux density for this source.  The overall structure is about 150 mas
($\sim 325$ pc) in size. A possible extension, not seen before, is visible in
our continuum image in the northern part of the source (see Fig.1).

\section{An off-nuclear \HI\ absorption}

The relatively narrow \HI\ absorption located at the systemic
velocity\footnote{We use the systemic velocity of V$_{\rm hel}$ = 36522\kms
($z = 0.12174 \pm 0.00002$) derived from the detailed analysis of the
emission lines presented in Holt et al.\ (2003).} was clearly detected.
Perhaps surprisingly, the absorption is not detected against the core or the
brighter southern radio lobe, but instead it is concentrated in a cloud-like
structure at the edge of the faint northern lobe.  In Fig.~1 the total
intensity image of the detected \HI\ absorption is shown superimposed on the
continuum emission.  The \HI\ cloud extends from about 20 mas ($\sim 43$ pc)
to $\sim 50$ mas ($\sim 108$ pc) from the nucleus.  It appears to be slightly
extended with an estimated size of $10 \times 30$ mas (corresponding to $22
\times 65$ pc). A small velocity gradient of $\lta 50$ \kms\ is observed
across the cloud as can be seen from the \HI\ profiles also shown in Fig.\ 1.
The FWHM of the
\HI\ profiles (at different positions as shown in Fig.\ 1) is about $\sim 150$
\kms.

Fig.\ 2 shows a comparison between the \HI\ profile derived from the
WSRT observations and the VLBI profile {\sl integrated} over the
cloud.  From the comparison it appears that, on the VLBI scale, we are
detecting basically the {\sl entire} narrow component of the \HI\ absorption
detected at lower resolution. 

The \HI\ absorption is located against the edge of the northern radio
lobe, against a region of very low brightness radio continuum.
The optical depth of the absorption is therefore extremely high.  The
peak of the \HI\ absorption is $\sim 20$ \mJybeam\ and the peak
optical depth reaches more than $60$\%.  The corresponding column
density of the \HI\ is up to $\sim 10^{22}\ T_{\rm spin}/(100\ {\rm
K})$ cm$^{-2}$.  This is likely to be a lower limit to the true column
density as the $T_{\rm spin}$ of a region so close to the AGN can be
as high as at least 1000~K (instead of 100~K which is more typical of
the cold, quiescent \HI\ in galaxy disks, Maloney et al. 1996).  Thus,
the true column density will likely be well in excess of $10^{22}$
cm$^{-2}$.

\section{Origin of the \HI\ absorption}

The intriguing result that comes out of these observations is that the
\HI\ absorption detected at the systemic velocity  is
originating from a cloud with high optical depth situated (in
projection) between 40 and 100 pc away from the core.
 
\HI\ detected at the systemic velocity is often considered to be 
associated with a circumnuclear disk or torus (see e.g. Conway \&
Blanco 1995, Peck \& Taylor 2001).  Given the observed characteristics
on the VLBI scale, this interpretation is not at all obvious in the
case of the \HI\ detected in 4C~12.50. This suggests that some care
should be taken when interpreting \HI\ absorption data from low
resolution observations, before the spatial information provided by
the VLBI is available.

The \HI\ absorption is located at the edge of the northern lobe/jet,
in a region of relatively weak radio continuum.  This suggests that
the medium producing the absorption is quite inhomogeneous with low
filling factor, since otherwise absorption would have been detected in
other locations.  The \HI\ column depth continues to increase up to
the edge of the radio source. This suggests that there may be more
atomic gas further to the south and/or west, where it can no longer be
seen in front of any radio emission. Thus, while the visible extent of
the absorber is about 22$\times$65 pc, its actual size could easily be
larger.

We find the properties of this absorber somewhat remarkable, and below
we describe several possible geometrical scenarios, none of which are
entirely conclusive.

\subsection{A circumnuclear structure}

A first possibility is that the visible absorbing region is indeed not
an isolated cloud, but forms part of a broader circumnuclear structure
around the centre, as expected from the traditional
orientation-unification models (see e.g.\ Antonucci 1993). This
structure occupies a roughly toroidal region around the centre,
through which the radio jets protrude into roughly the polar
directions. It could be that there is only a limited region, in the
northwestern sector, where there is both sufficient background radio
brightness and which has a large enough column depth in neutral gas to
obtain detectable \HI\ absorption against the counterjet. A modest
misalignment between the axes of the jets and the torus is then
indicated, but if the torus is geometrically quite thick, we could be
seeing only a region near its northern edge, while the atomic gas
density could increase towards the southwest, as surmised above, but
be invisible there for lack of radio background.

The entire southern radio structure could be protruding in front of the torus
in our direction, while at small (projected and physical) distances to the
core, the torus may no longer have a large enough column depth of atomic gas
(it could be ionised, see e.g the case of NGC~1052, Vermeulen et al. 2003). A
plausible projected position angle for the torus could be in the range $90\deg
- 120\deg$. We do not see the misalignment between this structure, on scales
of tens to a hundred pc, and the inner jets, which presumably are first
collimated by a central engine in an accretion disk geometry on much smaller
scales, as a fatal objection to the model. Indeed, whether or not jet engine
precession also plays a role (Lister et al.  2003), one could speculate that
the fact that the outer jets, on scales of tens to a hundred pc, are more
closely perpendicular to the surmised torus, is an indication of some bending
they undergo under the influence of the same density and pressure gradient of
which the torus is a manifestation.

However, the fact that we recover most of the flux detected at low resolution
indicates that the lack of more extended \HI\ absorption is not due to lack of
sensitivity. It is worth mentioning that in the other cases of
\HI\ absorption associated with a circumnuclear structure (e.g. 1946+708,
Cygnus~A, Hydra~A and 4C~31.04 studied by Peck \& Taylor 2001, Conway 1999,
Taylor 1996 and Conway 1997 respectively or even Arp 220, Mundell,
Ferruit \& Pedlar 2001) the source appears to be more homogenously covered by
the \HI\ screen instead of showing only one region of very high optical depth.

\subsection{An isolated \HI\ cloud: interaction or
chance alignment?}

We now consider the possibility that the detected \HI\ is coming from
an isolated cloud.  As mentioned above, the \HI\ cloud revealed by the
new VLBI observations is intriguingly located at the edge of the
northern lobe/jet.  This can be, of course, just a projection
effect. The transverse, projected distance of the absorber from the
centre is of order 50 pc but the cloud could be radially along our
line of sight at a much greater distance to the core, potentially even
at a kpc or more. However, it must then be part of a distribution of
clouds with a low enough number density and volume filling factor that
we see no other, potentially even a factor 40 lower column depth
clouds at similar distances in front of the larger, brigher southern
radio jet/lobe structure. In other words, the projection of this cloud
over the edge of the counterjet is fortuitous in this hypothesis.
However, it is also worth mentioning that, as noted before in the case
of other \HI\ absorption studies (see e.g.\ Peck \& Taylor 2001), the FWHM of
the detected \HI\ is significantly greater than that seen toward the
centre of our Galaxy ($\sim 20$ \kms, van der Hulst, Grolish \&
Haschick 1983) and so it seems unlikely that it is simply a cloud in
the host galaxy that happens to lie on the line of sight of the
central radio source.

Alternatively, the observed \HI\ absorber could have a line-of-sight
distance to the core similar to its projected distance, about 50
pc. Potentially, the radio counterjet could then be interacting with
the atomic gas. In this context, it appears remarkable that the cloud
is located in the region where the northern jet appears to slightly
bend and become less collimated.

A kinematical analysis has led Lister et al.\ (2003) to suggest that
4C~12.50 has a conical helical jet. This could arise from orbital
motion of the black-hole and/or precession of the jet nozzle. If this
is the case, the bending of the jet in the northern region could be
due to such a precession more than to some interaction with the
external medium.  Although one may wonder, given that the size of the
cloud is comparable to the projected distance to the centre, whether
such a cloud could be dynamicaly stable, the location
of the \HI\ cloud suggests that it is worth at least considering the
possibility of a physical association between the radio jet and the
\HI\ cloud suggesting that {\sl the cloud could be responsible for
bending the jet}.

On the other hand, an even sharper bend is observed in the southern
jet and one may wonder why no \HI\ absorption is detected there.
However, only gas that is {\sl in front} of the radio source is of
course detected via absorption. The geometry and the location of the
clumps is therefore extremely critical.

Regardless whether a direct interaction is actually occuring, the presence of
such a cloud suggests that the medium around the central regions of 4C~12.50
must be formed by dense and clumpy structures.  The presence of a clumpy
medium is also indicated by the polarization study.  Isolated regions of high
fractional radio polarization -- detected only when observed with the high
resolution of the VLBI -- were found by Lister et al. (2003) suggesting the
presence of a clumpy (and depolarizing) medium.

\begin{figure}
\centerline{\psfig{figure=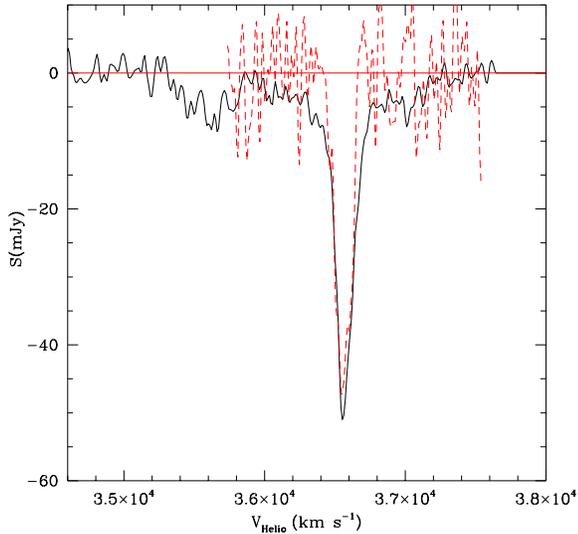,angle=0,width=7cm}}
\caption{Comparison between the WSRT profile (black) and the VLBI profile
integrated over the \HI\ absorption cloud (dashed).}
\end{figure}

\section{Effects of a clumpy medium}

In Sec. 4 we have shown that, although we cannot rule out completely other
possible explanations, a likely possibility is that we have detected an high
column density \HI\ cloud in the central region ($\sim 50$ pc) of 4C~12.50
against which the jet is interacting.

The presence of a medium formed by dense clumps surrounding the nuclear
regions can be expected, at least in the first stage of the radio galaxy's
evolution, as a result of a merging event. Clouds like the one detected in the
\HI\ observations of 4C~12.50 could represent the debris left over from the
merger that triggered the activity.  The presence of a clumpy medium can
therefore be a relatively common phenomena (see Sec. 6) that needs to be
taken into account in order to understand the evolution of the initial phase
of a newly-born (or restarted) radio jet.

In order to understand what would be the effect of this cloud on the jet, we
can estimate the mass of the \HI\ in the cloud.  As mentioned above, the
minimum estimated size is $22 \times 65$ pc.  Assuming an average column
density of $6 \times 10^{21}$ cm$^{-2}$ (which is a strict lower limit), we
estimate the \HI\ mass to be $\sim 7\times 10^4$ \msun.  As discussed above,
the column density could be easily a factor 10 higher, so this will give a few
times $10^5$ to $10^6$ \msun\ as \HI\ mass.  This is not too different from
the mass derived for the ionized gas (see Holt et al.\ 2003).  However, even
if a direct interaction occurs, the amount of gas does not seem to be enough
to frustrate the radio jet (see Carvalho 1994, 1998; one would need more than
$10^{10}$\msun).  It could, however, be able to {\sl momentarily} destroy the
path of the jet and therefore slow down the growth of the radio source until
the radio plasma clears its way out.  The similarities between the radio
morphology of 4C~12.50 and the simulation of a jet interacting with a clumpy
medium (Bicknell et al.  2003) are indirect support of this scenario.  In
these simulations, the jet is temporarely disrupted and decollimated due to
the interaction with a dense cloud in the ISM.  After a while this cloud is
destroyed and the jet collimates again and can continue in its expansion. 

Evidence that this could indeed be the situation is suggested also by the
study of the ionized gas (Holt et al.\ 2003).  In 4C~12.50 complex emission
line profiles are observed at the position of the (optical arcsec-scale)
nucleus. They have been interpreted as resulting from emission originating
from kinematically distinct regions with gradients in density, ionisation
potential and velocity.  This is what is expected if we are dealing with a young
radio galaxy with nuclear regions that are still enshrouded in a dense cocoon
of gas and dust and with the radio jets now expanding through this cocoon,
sweeping material out of the nuclear regions.  More examples of radio galaxies
in such a stage of their evolution do exist (see e.g.\ PKS 1549--79, Tadhunter
et al.\ 2001).

As mentioned in the introduction, the presence of a rich ISM in 4C~12.50 is
well known, e.g., from the high concentration of molecular gas found by Evans
et al.\ (1999) as well as the high far-IR luminosity and a young stellar
population component.  Because of this, 4C~12.50 has been often been suggested
as a prime candidate for the link between major-merger systems like the
ultraluminous infrared galaxies and radio galaxies.  \\ In the case of nearby
large mergers like the Antennae, molecular gas has been found in the form of
giant molecular clouds.  Following the work of e.g.\ Wilson et al.\ (2003),
those clouds can be relatively large (up to a few hundred pc in size), with a
$\Delta{\rm FWHM} \sim$ 20-50 \kms\ and a mass range between $5 \times 10^6 $
and $9 \times 10^8$ \msun.  The \HI\ cloud we detect is smaller than this and
less massive.  However, if the \HI\ corresponds only to that part of the gas
cloud that is photo-dissociated by the radiation from the AGN or the shocks
produced by the interaction with the radio jet, the total gas mass may fall in
the same range.  A major difference is the width of the line: the \HI\ cloud
in 4C~12.50 shows a much wider profile.  This could be an indication of a
kinematic disturbance due to the jet and further evidence that this gas is
indeed affected by some interaction with the radio jet. 

More gas clouds could be in the path of the radio plasma flow and, in
particular, some of them can be already ``run over'' by the jet. If that
happens, the clouds are compressed and they fragment. The fragments
can quickly radiate away their excess energy. This results in the formation of
dense and cool structures at high velocities (Mellema et al. 2002).  This part
of the gas can be responsible for the broad \HI\ absorption (detected in the
WSRT data), and be similar to the case of 3C~293 described in Morganti et
al. (2003c).

\section{A common phenomenon?}

4C~12.50 does not represent the first case where the presence of \HI\
absorption is associated with an off-nuclear cloud. One of the best
examples known is the Seyfert galaxy IC~5063 (Morganti et al.\ 1998,
2003b; Oosterloo et al.\ 2000) where the off-nuclear \HI\ absorption is
associated with a region of interaction between the radio jet and the
ISM (in particular a molecular cloud). In IC~5063, the presence of
interaction is further supported by the fact that the absorption is
highly blueshifted (up to 700 km/s), and therefore corresponds to a
fast outflow.

Among radio galaxies, 3C~236 shares many similarities with 4C~12.50. Although
it is a giant radio galaxy, it shows a restarted activity as indicated by the
nuclear component that has a structure similar to many CSS sources (Schilizzi
et al.\ 2001).  The VLBI study carried out by Conway \& Schilizzi (2000) has
shown that relatively narrow, but high opacity, absorption is associated with
the tip of the eastern jet (about 2~kpc away from the nucleus). Conway \&
Schilizzi (2000) explain this absorption as due to the photo-dissociation of
molecular gas by the radiation from the shock produced by interaction between
the radio jet and a molecular cloud.  Further down along the lobe, a broader
component of the \HI\ absorption is seen, indicating (according to Conway \&
Schilizzi 2000) that those regions have been reached and disturbed by the
shock.

Another, more recent, example of off-nuclear \HI\ absorption has been found
in the Compact Symmetric object 4C~37.11 (or 0402+379, see Maness et al.\ 2004)
where broad ($\sim 500$ \kms, but in this case redshifted compared to the
systemic) absorption line was found in the region of the southern hot-spot.

\section{Conclusions}

VLBI observations of neutral hydrogen in the radio galaxy 4C~12.50 have given
that perhaps surprising result that most of the \HI\ detected close to the
systemic velocity is concentrated in a high column density cloud situated - in
projection - at about 50~pc from the nucleus.

The radio activity in 4C~12.50 is considered to be triggered by the
merger of gas-rich galaxies for a number of reasons (e.g. the large
amount of molecular gas detected, by Evans et al. 1999). This is also
confirmed by the study of the stellar population of the host galaxy
(Tadhunter et al. 2004), where a substantial contribution from a young
stellar component is found, implying a major merger as origin of this
system. The young stellar component is, however, relatively old
(between 0.5 and 1.5 Gyr) suggesting that the activity was triggered
late in the merger process. 

The \HI\ cloud detected in 4C~12.50 indicates, indeed, that, long
after the merger took place, a  clumpy interstellar medium is
present in the nuclear regions of the galaxy.   The
presence of such a clumpy medium supports the idea (also suggested by
the observations of ionized gas) that the radio jets may experience
strong interaction with this medium before the radio plasma
clears its way out.  Clouds like the one detected in 4C~12.50 are
likely to be responsable for the bending and decollimation of the
young radio jet.  A similar distribution of the ISM may be common in
radio galaxies for which the radio activity just started (or
restarted, as it is the case in 3C~236).

Finally, the VLBI observations of the neutral hydrogen in 4C~12.50 have shown
that \HI\ absorption detected close to the systemic velocity in many radio
galaxies cannot be taken (by itself) as evidence for being associated with
circumnuclear tori or disks, but high resolution follow up are always
necessary to get better understanding of the actual origin of this gas.

\end{document}